# Semantic Integration in the IFF

Robert E. Kent

*Ontologos*

rekent@ontologos.org

## Abstract

*The IEEE P1600.1 Standard Upper Ontology (SUO) project aims to specify an upper ontology that will provide a structure and a set of general concepts upon which domain ontologies could be constructed. The Information Flow Framework (IFF), which is being developed under the auspices of the SUO Working Group, represents the structural aspect of the SUO. The IFF is based on category theory. Semantic integration of object-level ontologies in the IFF is represented with its fusion construction[*]. The IFF maintains ontologies using powerful composition primitives, which includes the fusion construction.*

## 1. The Information Flow Framework

The IEEE P1600.1 Standard Upper Ontology (SUO)[1] project aims to specify an upper ontology that will provide a structure and a set of general concepts upon which object-level domain ontologies could be constructed. These object-level domain ontologies will utilize the SUO for "applications such as data interoperability, information search and retrieval, automated inferencing, and natural language processing". A central purpose of the SUO project is interoperability.

The Information Flow Framework (IFF)[2] is being developed to represent the structural aspect of the SUO. It aims to provide semantic interoperability among various object-level ontologies. The IFF supports this interoperability by its architecture and its use of a particular branch of mathematics known as category theory (Mac Lane, 1971). A major reason that the IFF uses the architecture and formalisms that it does is to support modular ontology development. Modularity facilitates the development, testing, maintenance, and use of ontologies. The categorical approach of the IFF provides a principled framework for modular design via a structural metatheory of object-level ontologies. Such a metatheory is a method for representing the structural relationships between ontologies.

The IFF provides mechanisms for the principled foundation of a metalevel ontological framework – a framework for sharing ontologies, manipulating ontologies as objects, relating ontologies through morphisms, partitioning ontologies, composing ontologies via fusions, noting dependencies between ontologies, declaring the use of other ontologies[3], etc. The IFF takes a building blocks approach towards the development of object-level ontological structure. This is a rather elaborate categorical approach, which uses insights and ideas from the theory of distributed logic known as information flow (Barwise and Seligman, 1997) and the theory of formal concept analysis (Ganter and Wille, 1999). The IFF represents metalogic, and as such operates at the structural level of ontologies. In the IFF, there is a precise boundary between the metalevel and the object level.

The modular architecture of the IFF consists of metalevels, namespaces and meta-ontologies. There are three metalevels: top, upper and lower. This partition, which corresponds to the set-theoretic distinction between small (sets), large (classes) and generic collections, is permanent. Each metalevel services the level below by providing a language that is used to declare and axiomatize that level. The top metalevel services the upper metalevel, the upper metalevel services the lower metalevel, and the lower metalevel services the object-level. Within each metalevel, the terminology is partitioned into namespaces[†]. The number of namespaces and the content may vary over time: new namespaces may be created or old namespaces may be deprecated, and new terminology and axiomatization within any particular namespace may change. In addition, within each level, various namespaces are collected together into meaningful composites called meta-ontologies. At any particular metalevel, these meta-ontologies cover all the namespaces at that level, but they may overlap. The number of meta-ontologies and the content of any meta-ontology may vary over time: new meta-ontologies may be created or old meta-ontologies may be deprecated, and new namespaces within any particular meta-ontology may change (new versions).

The top IFF metalevel provides an interface between the simple IFF-KIF language and the other IFF terminology. By analogy, the simple IFF-KIF language is like a machine language and the top IFF metalevel is like an assembly language. There is only one namespace and one meta-ontology in the top metalevel: the Top Core (meta) Ontology. This meta-ontology represents generic collections. In a sense, it bootstraps the rest of the IFF into existence. The single namespace, the meta-ontology and the top metalevel can be identified with each other. The upper and lower IFF metalevels represent the structural aspect of the SUO. By analogy, the structural aspect of the SUO is like a high lev-

---

[*] Throughout this paper, we use the intuitive terminology of mathematical context, passage/construction, pair of invertible passages and fusion for the mathematical concepts of category, functor, adjunction and colimit, respectively.

[†] The IFF terminology is disambiguated via the disjoint union of local namespace terminology. A fully qualified term in the IFF is of the form "ν$τ", where the namespace prefix label "ν" is a "." separated sequence of alphabetic strings that uniquely represents an IFF namespace, and the local unqualified term "τ" is a unique lowercase alphanumeric-dash string within that namespace. For example: the term

        `th.col.psh$coequalizer-diagram`

represents the coequalizer diagram underlying a pushout diagram of theories within the theory pushout namespace in the lower IFF metalevel.

el programming language such as Lisp, Java, ML, etc. There are three permanent meta-ontologies in the upper metalevel: the Upper Core (meta) Ontology represents the large collections called classes; the Category Theory (meta) Ontology represents category theory; and the Upper Classification (meta) Ontology represents information flow and formal concept analysis. There will eventually be many meta-ontologies situated in the lower IFF metalevel[‡]. Currently there are only four: the Lower Core (meta) Ontology represents the small collections called sets; the Lower Classification (meta) Ontology is a small and more specialized version of its upper counterpart; the Algebraic Theory (meta) Ontology represents equational logic; and the Ontology (meta) Ontology represents first order logic and model theory. All versions of these meta-ontologies are listed as links in the SUO IFF site map[4].

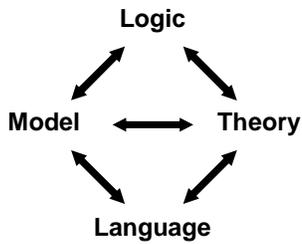

**Figure 1: IFF-OO Architecture**

The IFF, which is situated at the metalevel, represents form. The ontologies, which are situated at the object level, represent content[§]. By analogy, the content aspect of the SUO is like the various software applications, such as word processors, browsers, spreadsheet software, databases, etc. The distinction between content and form is basic in the general grammar of natural languages, in logic and in ontology. In all of these realms, but especially in logic and ontology, the IFF offers a coherent principled approach to form. Such form is realized in the structuring, mapping and integration of ontologies. The IFF offers axiomatization and techniques for the hierarchical structuring of object-level ontologies via the lattice of theories, the mapping between ontologies via syntax directed translation, and the semantic integration of ontologies via mediating or reference ontologies. To paraphrase John Sowa[5], developing the tools and methodologies for extending, refining, and sharing object-level ontologies is more important than developing the content for those ontologies.

## 2. Basic Concepts of the IFF-OO

The metalevel axiomatic framework for object-level ontologies represented in first order logic and model theory is concentrated in the lower metalevel IFF Ontology (meta) Ontology (IFF-OO). The IFF-OO is a generic framework for the representation and manipulation of object-level ontologies. The architecture of the IFF-OO (Figure 1) consists of four central mathematical contexts[*] interconnected by five pairs of invertible passages[*]. Each of the four contexts represents a basic concept axiomatized in the IFF-OO. These four concepts are language, theory, model and logic. The context of first order logic *languages*[6] sits at the base of the IFF-OO – everything depends upon it. The three other contexts – models, theories and logics – are situated above the language context. Models provide the interpretive semantics for object-level ontologies, *theories* provide the formal or axiomatic semantics, and logics provide the combined semantics. Any theory is based on a language, and the context of theories is connected to the context of languages by the base passage. An object-level ontology is populated when it has instance data. Unpopulated object-level ontologies are represented by IFF theories, whereas populated object-level ontologies are represented by IFF logics. This paper deals only with formal, axiomatic semantics for object-level ontologies. Interpretive semantics will be combined with this in future work.

The concept of an IFF language is many-sorted – the definition follows (Enderton, 1972), generalizing the standard notion of a single-sorted language. The IFF terminology is somewhat different from Enderton – it uses the two polarities of entities versus relations and instances versus types: an IFF entity type corresponds to a sort, an IFF relation type corresponds to a predicate, and an IFF function type corresponds to a function symbol. In this paper, we ignore function types for simplicity – these are adequately handled in the IFF Algebraic Theory (meta) Ontology. Note that an IFF language deals only with type information. Constants are regarded as nullary function types. Languages are comparable via *language morphisms*, and theories are comparable via *theory morphisms*. Any language $L$ determines a *lattice of theories* $fiber(L)$[**], a base passage $fiber$[††]. Any language morphism $f: L_1 \to L_2$ determines a function $expr(f): expr(L_1) \to expr(L_2)$ by induction, and from this a *lattice morphism of theories*

$$fiber(f) = \langle inv(f), dir^\exists \rangle : fiber(L_2) \to fiber(L_1),$$

---

[‡] A module in the IFF lower metalevel should represent a well-researched area. In addition to the IFF-OO, which represents first order logic and model theory, other non-core lower metalevel modules are also being considered: a module for the "soft computation" of both rough sets and fuzzy logic; a module for theories of semiotics; a module for game-theoretic semantics; etc.

[§] Many current object-level ontologies contain generic axiomatizations for notions such as binary relations, partial orders, etc. In the IFF, these are not needed, since such axiomatizations are included in the Lower Core (meta) Ontology, etc. When compliant with the IFF, object-level ontologies can concentrate on their core axiomatics.

[**] The *lattice of theories* $fiber(L)$ for a language $L$ is the complete lattice of all theories with base language $L$ using entailment order between theories: $T_2 \le T_1$ means that $T_2$ is more specialized than $T_1$ in the sense that $T_1$ is contained in the closure of $T_2$; or equivalently, that any theorem of $T_1$ is entailed by the axioms of $T_2$.

[††] A *fiber* of a passage $P: \mathbf{C} \to \mathbf{B}$ for fixed object $b \in \mathbf{B}$ is analogous to the inverse image of $b$ along $P$, thus forming the sub-context $fiber_P(b) \subseteq \mathbf{C}$ of all $\mathbf{C}$-objects that map to $b$ and all $\mathbf{C}$-morphisms that map to the identity at $b$.

the fiber invertible passages of direct/inverse image operators – the (*existential*) *direct image* operator

$$dir^\exists(f) = (\exists expr(f))^{op} : fiber(L_1) \to fiber(L_2)^{\ddagger\ddagger}$$

and the *inverse image* operator

$$inv(f) = (expr(f)^{-1})^{op} : fiber(L_2) \to fiber(L_1).$$

The mapping of unpopulated object-level ontologies is represented by IFF language/theory morphisms. In particular, the IFF represents ontology mapping as the movement of theories back and forth between lattices of theories by using the above lattice morphism of theories over a language morphism.

A recent vote by the SUO Working Group approved a proposal by John Sowa to develop a library of modules structured in a hierarchy. This library of modules will include modules derived from other object-level ontologies. The hierarchical structure framing such a library of modules is a lattice of theories. Sowa has offered a step-wise approach for building a library of modules[7]. However, the processing involved here can be applied to any system of ontologies, and each step of Sowa's process of "building the hierarchy" is represented in the IFF. To do this we represent a module as an IFF theory. A library of modules, regarded as a generalization-specialization hierarchy, is conceptually situated within the context of a lattice of theories[**] and its correlated structure known as the *truth concept lattice*[§§]. In the IFF, an unpopulated <u>monolithic</u> object-level ontology is represented as an IFF theory, the same as a module. The IFF regards a library of modules to be an unpopulated <u>modularized</u> object-level ontology. This is represented in the IFF as a *diagram of theories*[***]. In other terminology, an IFF diagram of theories represents a system of object-level ontologies. Diagrams of theories are comparable via *theory diagram morphisms*[6]. Any diagram of theories $T$ indexed by a shape graph $G$ has a base diagram of languages $L = base(T)$ of the same shape, where the language (language morphism) at any indexing node (edge) of graph $G$ is the underlying base language (language morphism) of the theory (theory morphism) at that node (edge). Generalizing the fiber over a language, any language diagram $L : G \to |\text{Language}|$ determines a *lattice of theory diagrams fiber(L)*[6]. Generalizing the fiber adjoint pair over a language morphism, any language diagram morphism φ determines a *lattice morphism of theory diagrams fiber(φ)*[6].

## 3. Fusion of a System of Ontologies

The IFF can utilize the fusion construction[*] in various mathematical contexts. Since this paper only discusses the formal, axiomatic semantics of integration, here we limit ourselves to the fusion construction for languages and theories. The fusion of theories is defined in terms of the fusion of languages (Table 1).

### Table 1: The Fusion Construction[†††]

1. Informally, identify the theories to be used in the construction.
2. Formally, create a diagram of theories $T$ of shape (indexing) graph $G$ that indicates this selection. This diagram of theories is transient, since it will be used only for this computation. Other diagrams could be used for other fusion constructions.
3. Form the fusion theory $T^\cdot = \sum T$ of this diagram of theories, with theory fusion cocone $\tau : T \Rightarrow T^\cdot$.
   a. Compute the base diagram of languages $L = base(T)$ with the same shape. In more detail, $L = base(T)$
      $= \{L_n\} + \{L_e : L_m \to L_n\}$
      $= \{base(T_n)\} + \{base(T_e) : base(T_m) \to base(T_n)\}$.
   b. Form the fusion language $\boldsymbol{L} = \sum L$ of this diagram, with language fusion cocone $\lambda : L \Rightarrow \boldsymbol{L}$. In more detail, $\lambda = \{\lambda_n : L_n \to \boldsymbol{L}\}$, satisfying the conditions $\lambda_m = L_e \cdot \lambda_n$ for $G$-edge $e : m \to n$.
   c. Move (the individual theories $\{T_n\}$ in) the diagram of theories $T$ from the lattice of theory diagrams $fiber(L)$ along the language morphisms in the fusion cocone $\lambda : L \Rightarrow \boldsymbol{L}$ to the lattice of theories $fiber(\boldsymbol{L})$ using the direct image function, getting the homogeneous diagram of theories $dir(\lambda)(T)$ with the same shape $G$, where each theory $dir(\lambda)(T)_n = dir(\lambda_n)(T_n)$ has the same base language $\boldsymbol{L}$ (the meaning of homogeneous).
   d. Compute the meet (union) of the diagram $dir(\lambda)(T)$ within the lattice $fiber(\boldsymbol{L})$ getting the fusion theory $T^\cdot = \sum T = meet(\boldsymbol{L})(dir(\lambda)(T))$.
   e. The language fusion cocone is the base of the theory fusion cocone $\lambda = base(\tau) : base(T) \Rightarrow base(T^\cdot)$.

As mentioned before, any diagram of theories $T$ has a base diagram of languages $L = base(T)$ of the same shape. It is important to note that the indexed theories within $T$ do not necessarily have the same base language. To semantically compare these theories and to conceptually situate them within a lattice of theories, we move them to the lattice of theories over the fusion language $\boldsymbol{L} = \sum L$, with this movement guided along the language morphisms in the fusion cocone $\lambda : L \Rightarrow \boldsymbol{L}$. The latter is a $node(G)$-indexed collection of language morphisms, whose source is the language diagram $L$ and whose target is the language $\boldsymbol{L}$. For any diagram of theories $T$ in $fiber(L)$, the *direct image fiber* operator $dir(\lambda)$ moves $T$ along the fusion cocone to $dir(\lambda)(T)$, a homogeneous diagram of shape $G$ in the lattice of theories over $L$. Homogeneous means that all the indexed theories in $dir(\lambda)(T)$ have the same base language $L$, and hence can be semantically compared via the theory entailment order. The fusion of the diagram of theories $T$ resolves into $\Sigma T = meet(L)(dir(\lambda)(T))$ – the fiber direct image $dir(\lambda)$ along the base diagram fusion cocone, followed by the meet $meet(L)$ in the lattice of theories over $L$, the base diagram fusion language.

---

[‡‡] In the following, we abbreviate this as $dir(f) = dir^\exists(f)$.

[§§] Intuitively, the truth concept lattice is the lattice of closed theories. The lattice order is reverse subset inclusion. The truth concept lattice is the concept lattice for the *truth classification*, the fundamental example 4.6 introduced in (Barwise and Seligman, 1997).

[***] A *diagram of theories* $T : G \to |\text{Theory}|$ consists of two collections, theories and theory morphisms, indexed by a *shape* graph $G$: each $G$-node $n$ indexes a theory $T_n$ and each $G$-edge $e : m \to n$ indexes a theory morphism $T_e : T_m \to T_n$. The size of a diagram corresponds to the cardinality of the node and edge sets of its shape graph. Although these can be infinite, in most practical situations they are finite – there are empty diagrams, single theory diagrams, diagrams with only two theories and one theory morphism, etc.

[†††] The two operations of (1) forming *sums* of theories and (2) specifying *endorelations* and then computing their *quotients*, offer an alternate method for the fusion construction of diagrams of theories: *coequalizers* of theories can be constructed as quotients of endorelations; and *pushouts* of theories can be constructed in terms of sums of components and then quotients of endorelations.

Two new ideas have emerged recently in the discussion of the SUO Working Group: the idea of a polycosmos and the idea of mapping closure. Both of these ideas are important in the theory of semantic integration. However, it was not possible to succinctly express these ideas without the use of theory fusions.

○ The idea of a *polycosmos*[‡‡‡] was first expressed[8] by Patrick Cassidy: a polycosmos is an unpopulated modular object-level "ontology that has a provision for alternative possible worlds, and includes some alternative logically contradictory theories as applying to alternative possible worlds". The mathematical formulation of polycosmic[9] was immediately given by the author in terms of the fusion of a diagram of theories. A diagram of theories $T$ is *monocosmic* when the fusion theory $\Sigma T$ is consistent. A diagram of theories $T$ is *pointwise consistent* when each indexed theory in $dir(\lambda)(T)$ is consistent. A monocosmic diagram of theories is pointwise consistent by default. A diagram of theories $T$ is *polycosmic* when it is pointwise consistent, but not monocosmic; that is, when there are (at least) two consistent but mutually inconsistent theories in $dir(\lambda)(T)$. In the IFF[§§§], there are some extreme polycosmic diagrams of theories, where any two theories are either equivalent or mutually inconsistent. Each of the theories in these diagrams lies at the lowest level in the lattice of theories, strictly above the bottom inconsistent theory containing all expressions.

○ The idea of *mapping closure* was first expressed[10] by the author. Any mapping of ontologies involves this notion of mapping closure. For any morphism of languages $f: L_1 \to L_2$, the mapping closure of $f$ applied to any source theory $T_1 \in fiber(L_1)$ is the closure associated with the fiber adjoint pair: $clo(f)(T_1) = inv(f)(dir(f)(T_1))$. Since language morphisms and endorelations are in a sense equivalent[****], the idea of mapping closure is also induced by a language endorelation. An endorelation based on a language $L$ defines by induction an equivalence relation on variables, entity types, relation types and expressions. One expression is equivalent to another expression when the constituent terms[††††] in each are equivalent. Any expression that is equivalent to a theorem of a theory $T \in fiber(L)$ is included in the mapping closure[‡‡‡‡].

Any morphism of languages $f: L_1 \to L_2$ determines a *lattice morphism of theories*

$$\langle dir^\forall(f), inv(f) \rangle : fiber(L_1) \to fiber(L_2)$$

with the (*universal*) *direct image* operator

$$dir^\forall(f) = \forall expr(f)^{op} : fiber(L_1) \to fiber(L_2)$$

and the *inverse image* operator. In summary, for any morphism of languages $f: L_1 \to L_2$ there are two linked pairs of invertible monotonic functions:

$$dir^\forall(f) \dashv inv(f) \dashv dir^\exists,$$

with $dir^\forall(f)$ and $inv(f)$ preserving joins (intersections), and $inv(f)$ and $dir(f) = dir^\exists$ preserving meets (unions). Two questions arise. (1) What is the significance of the mapping closure? (2) Which quantificational direct image operator should be used for moving theories? In the IFF view, mapping theories along a language morphism requires a commitment to mapping closure. In other words, if one is willing to use a language morphism to map a theory, then one is committing oneself to the mapping closure of that theory; that is, one is essentially asserting all of the additional axioms in the difference between the theory and its mapping closure. The existential direct image operator is seen to be important by its use in the fusion construction. However, what about the universal direct image operator? The fact is that the two operators are identical on the mapping closure of a theory. Hence, if we commit ourselves to the mapping closure of a theory, it does not matter which direct image operator we use, since they are both equal in this case.

## 4. Maintenance of a System of Ontologies

This section discusses how the notions of modularity and centralization are represented in the IFF. As the author has discussed[11] and demonstrated[12], each step of Sowa's process of "building the hierarchy"[7] is represented in the IFF. All steps take place in the context of theories. However, in the general maintenance of a diagram of theories, these processing steps can be used in any fashion deemed necessary. The following are various operations that are

---

[‡‡‡] According to the dictionary, a cosmos is an orderly harmonious systematic universe.

[§§§] Since IFF models have a set of tuples (= relation instances) as one component, they are more refined than traditional model-theoretic structures and are better able to represent the intuitive notion of context – some IFF models even have only one tuple.

[****] Any language morphism has a kernel (equivalence) endorelation based on the source language, where two source types are equivalent when they are mapped to the same target type. Conversely, any language endorelation generates an epimorphic language morphism onto the quotient language of the endorelation.

[††††] By terms, we mean the variables and the entity, relation and function types used in the language $L$. Constants are nullary function symbols.

[‡‡‡‡] The IFF notion of language endorelation is a theory of relative synonymy – synonymy relative to the base language, and hence relative to the conceptual structures of whatever community owns and manages the corresponding ontology. Such a theory of relative synonymy may be related to any linguistic/philosophical discussion of synonymy, such as (Quine, 1951).

possible in the IFF in order to practically maintain a diagram of theories.
- **Consistency checking:** Any theory in a homogeneous diagram of theories may be inconsistent (equivalent to the bottom of the lattice of theories). A basic and non-trivial operation is to check for the consistency of the indexed theories in a diagram. Of course, any theory that comes with its own special model is already consistent.
- **Sum theory:** This is a procedure for distinguishing the various terms used in a discrete diagram of theories. Every theory in such a diagram has a unique theory index, and all terms in the standard theory sum are distinguished by 'labeling'' with the index of their theory of origin. This is the process of forming the sum in the context of theories and the underlying context of languages.
- **Endorelation and Quotient theory:** The quotient of a theory is based upon an endorelation over that theory[§§§§]. The identification of pairs of terms[††††] corresponds to the mathematical process of forming the *quotient* of the sets of terms in a theory via a suitable *endorelation*. This is the process of forming the quotient in the context of theories and the underlying context of type languages.
- **Subtheory:** Often it is helpful in maintaining a diagram of theories to extract smaller (and hence more generic) subtheories from larger more specific ones. This makes the diagram of theories more flexible to use. In particular, when fusing theories, one may need to only use some smaller more generic parts. Each extracted theory is more general than its theory of origin, and thus higher in the lattice hierarchy.
- **Alignment:** For alignment in particular and integration in general, we follow the definitions of the ontology working group of the NCITS T2 Committee on Information Interchange and Interpretation as recorded by Sowa[3]. Ontological alignment consists of the sharing of common terminology and semantics through a mediating or reference ontology (Kent, 2000). The intent of alignment is that mapped types are equivalent. Such equivalence can be automatically computed via the FCA-Merge process (Stumme and Mädche, 2001)[*****]. To formalize this, we represent an equivalence pair of types as a single type in a mediating or reference theory, with two mappings from this new type back to the participant theory types. Thus, alignment is represented as a span or 'Λ'-shaped diagram of three theories and two theory morphisms. The mediating or reference ontology in the middle represents both the equivalenced types and the axiomatization needed for the desired degree of compatibility with the participant ontologies, whether partial or complete. Since the theoretical alignment links preserve this axiomatization, compatibility will be enforced[†††††].
- **Sum diagram:** Given two diagrams of theories $T_1$ and $T_2$ of shapes $G_1$ and $G_2$, respectively, the sum diagram of theories $T = [T_1, T_2]$ has the sum shape $G_1 + G_2$ with object function $obj(T)$ that maps nodes in $node(G_1 + G_2) = node(G_1) + node(G_2)$ according to component: $obj(T)(n_1) = obj(T_1)(n_1)$ and $obj(T)(n_2) = obj(T_2)(n_2)$; similarly for edges.
- **Removal:** Any theory in a diagram might be mark for deletion for various reasons – the theory may have been proven inconsistent, or the theory may no longer be of interest to the community federation maintaining the system of ontologies.
- **Fusion (or Unification):** It may be desirable at any time to create a customized theory. One example of such a customized theory is a "great big hierarchy with modules copied in, frozen into place, and relabeled to avoid inconsistencies" as described[13] by John Sowa. This is built as the fusion construction of a sub-diagram of theories (Table 1). The fusion $T^\bullet$, the desired theory to be constructed, is just another theory. The other theories in the diagram being maintained have been left in place undisturbed. Forming the meet is a special case of the fusion construction for a homogeneous sub-diagram of theories.
- **Theory Creation:** Often a small theory of specialized axioms is needed. This may occur when defining a customized theory as the fusion of a diagram with the small theory as one indexed component.

## 5. Future Prospects

Full semantic integration involves the notion of information flow (Barwise and Seligman, 1997). Special cases of this have appeared in the papers (Kent, 2000 and 2003), (Kalfoglou and Schorlemmer, 2002) and (Schorlemmer and Kalfoglou, 2003). In particular, the papers by the author argue that the semantic integration of ontologies is the two-step process of alignment and unification. Ontological

---

[§§§§] This is a systematic procedure for specifying the pairs of terms to be semantically identified. One can assume that the terms in the sum of a (discrete) diagram of theories are coordinated with one another in the following sense. In a theory sum, (1) any two terms from independently developed component theories should not be identified; however, (2) two identical terms from different component theories should be identified if these theories originated by subsetting from a third more specialized theory.

[*****] In fact, although we recognize that it can serendipitously discover new relationships, we view FCA-Merge as predominately an automatic process for ontology alignment. It is important to note that FCA-Merge requires interpretative or combined semantics, since it crucially depends upon instance data and classifications. Hence, this approach to alignment uses logics, not just theories.

[†††††] In general, alignment acts through community ontology port(al)s. Before two ontologies can be aligned, it may be necessary to introduce new subtypes or supertypes of terms in either ontology in order to provide suitable targets for alignment. In addition, when any participant ontology has some distinct instance data, alignment may quotient that participant. Hence, alignment is represented by a 'W'-shaped diagram, with the original participating ontologies at the two upper outer vertices, the mediating or reference ontology at the upper center vertex, and the participant port(al) ontologies at the two lower vertices.

alignment consists of the sharing of common terminology and semantics through a mediating or reference ontology. Ontological unification, concentrated in a virtual ontology of community connections, is fusion of the alignment diagram of participant community ontologies – the quotient of the sum of the participant port(al)s modulo the ontological alignment structure. The current paper contributes to this "information flow approach to semantic integration" by describing how the IFF represents formal semantic integration through its general fusion construction and situates formal semantic integration in the on-going maintenance of a system of ontologies. However, true information flow, and hence combined semantic integration, both formal and interpretive, occurs at the level of logics. The correct formulation of this requires the notion of free logics and the notion of fusions of logics. The current version of the IFF-OO has axiomatizations for free logics and for fusions of theories. However, fusions of logics cannot be constructed. The problem is that the current version of the IFF-OO follows too closely Enderton's notion of a sorted language. In particular, IFF languages using reference functions (sort functions in Enderton's terminology) cause problems when trying to construct the coproduct of models or logics. This has been remedied in the new version of the IFF-OO to be posted soon. See the discussion of the "IFF Work in Progress"[14] for more on this.

In summary, we argue that the principled framework of the IFF realizes the information flow approach to semantic integration, and we hope that this theoretical approach and its implementation[‡‡‡‡‡] will contribute to realization of the "gold standard for semantic integration" (Uschold and Gruninger, 2002).

---

[‡‡‡‡‡] The IFF takes the high road to implementation. There is work in progress on an IFF representation of the Meta Object Facility (MOF) and the Model Driven Architecture (MDA) of the Object Management Group (OMG). In the other direction, there are on-going explorations to demonstrate how the MOF can be used for a high level specification of the IFF.

[1] "IEEE P1600.1 Standard Upper Ontology (SUO) Working Group". http://suo.ieee.org/.

[2] "The SUO Information Flow Framework (SUO IFF)". http://suo.ieee.org/IFF/.

[3] Sowa, John F. Building, Sharing, and Merging Ontologies. Unpublished paper. (2001) http://www.jfsowa.com/ontology/ontoshar.htm.

[4] "The SUO IF Site Map". http://suo.ieee.org/IFF/site-map.html.

[5] Sowa, John F. "Reply: Ontology Structure & Content". (13 January 2001) http://grouper.ieee.org/groups/suo/email/msg02765.html.

[6] Kent, Robert E. "IFF Lattice of Theories (LOT) Glossary". http://suo.ieee.org/IFF/work-in-progress/LOT/glossary.pdf.

[7] Sowa, John F. "Building the hierarchy". (16 May 2003) http://suo.ieee.org/email/msg09453.html.

[8] Cassidy, Patrick. "Reply: "SUO Ballot with 2 Questions – 'monolithic'?". (10 Jun 2003) http://grouper.ieee.org/groups/suo/email/msg09701.html.

[9] Kent, Robert E. "Mathematical Definitions for Monocosmic and Polycosmic". (14 Jun 2003) http://grouper.ieee.org/groups/suo/email/msg09768.html.

[10] Kent, Robert E. "Mapping Closure". (5 Jul 2003) http://grouper.ieee.org/groups/suo/email/msg10356.html.

[11] Kent, Robert E. "Building the Hierarchy by Language and Module Processing". (2 Jun 2003) http://grouper.ieee.org/groups/suo/email/msg09555.html.

[12] Kent, Robert E. "Language and Module Processing" (31 May 2003) http://suo.ieee.org/IFF/work-in-progress/LOT/language-module-processing.html.

[13] Sowa, John F. "Reply: Charter vs. Consensus". (27 Jun 2003) http://grouper.ieee.org/groups/suo/email/msg10139.html.

[14] Kent, Robert E. "IFF Work in Progress". (4 April 2003) http://suo.ieee.org/IFF/work-in-progress/.